\pdfoutput=1

\documentclass[11pt]{article}

\usepackage[preprint]{nlp4MusA}

\usepackage{times}
\usepackage{latexsym}
\usepackage{amssymb,amsmath,amsfonts}
\usepackage{lipsum}
\usepackage{url}
\usepackage{pgfplots}
\usepgfplotslibrary{fillbetween}
\usepgfplotslibrary{external}
\usepgfplotslibrary{groupplots}
\usepackage[T1]{fontenc}

\usepackage[utf8]{inputenc}

\usepackage{microtype}

\usepackage{inconsolata}

\usepackage{subcaption}
\usepackage{graphicx}

\usepackage{mdframed}
\newmdenv[
  topline=false,
  bottomline=false,
  rightline=false,
  linecolor=purple,
  linewidth=2pt
]{rewrite}

\newcommand{\later}[1]{}

\newcommand{\tableSection}[1]{}

\newcommand{\availablemodel}[1]{\href{#1}{\ding{51}}}
\newcommand{\notavailablemodel}[1]{\ding{55}}

\title{Analyzing Byte-Pair Encoding on Monophonic and Polyphonic\\Symbolic Music: A Focus on Musical Phrase Segmentation}

\author{
 \textbf{Dinh-Viet-Toan Le\textsuperscript{1}},
 \textbf{Louis Bigo\textsuperscript{2}} \and
 \textbf{Mikaela Keller\textsuperscript{1}}
\\
\\
 \textsuperscript{1}Univ. Lille, CNRS, Inria, Centrale Lille, UMR 9189 CRIStAL, F-59000 Lille \\
 \textsuperscript{2}Univ. Bordeaux, CNRS, Bordeaux INP, LaBRI, UMR 5800, F-33400 Talence
\\
 \small{\texttt{dinhviettoan.le@univ-lille.fr}}
}

\begin{document}
\maketitle

\begin{abstract}
Byte-Pair Encoding (BPE) is an algorithm commonly used in Natural Language Processing to build a vocabulary of subwords, which has been recently applied to symbolic music.
Given that symbolic music can differ significantly from text, particularly with polyphony, we investigate how BPE behaves with different types of musical content. 
This study provides a qualitative analysis of BPE's behavior across various instrumentations and evaluates its impact on a musical phrase segmentation task for both monophonic and polyphonic music. 
Our findings show that the BPE training process is highly dependent on the instrumentation and that BPE ``supertokens'' succeed in capturing 
abstract musical content. 
In a musical phrase segmentation task, BPE notably improves performance in a polyphonic setting, but enhances performance in monophonic tunes only within a specific range of BPE merges.
\end{abstract}

\section{Introduction}
\label{sec:intro}

A major similarity between text and music lies in their nature as semiotic systems, 
as they can be represented as sequences of elements~\cite{lerdahl2013musical}.
This common characteristic has led to numerous adaptations of Natural Language Processing (NLP) methods in the domain of symbolic music analysis and generation~\cite{le2024natural}. 
Mirroring the view of a text as a sequence of tokens representing words, subwords or characters, \textit{tokenization} practices have also been adopted to process symbolic music. Several choices of types of musical "characters" and various tokenization algorithms to segment the sequence of musical "characters" have been proposed~\cite{kumar2023words}.

However, music profoundly differs from text, notably because of some structural characteristics such as rhythm or polyphony~\cite{jackendoff2009parallels}. We can, thus expect tokenization algorithms such as Byte-Pair Encoding (BPE)~\cite{sennrich2016neural} to behave differently when applied to text or music.
The aim of this study is to highlight some commonalities and differences in BPE behaviors with multiple types of music as compared to text.
This work is twofold: we first propose a statistical description of the vocabulary of tokens obtained when BPE is applied to text compared to the vocabularies obtained with various types of music. This comparison highlights some musical properties captured by this tokenization algorithm (Section~\ref{sec:analyzing}).
Informed by these observations, we then focus on a downstream task, musical phrase segmentation, to quantitatively compare the impact of BPE on monophonic and polyphonic music (Section~\ref{sec:experiments}).

\section{Subword tokenization in symbolic music}
\label{sec:subword_tok}

Subword tokenization, where tokens are subwords instead of characters or words, is a common practice in NLP. 
It is used to deal with out-of-vocabulary words that are obtained by combining multiple subwords.
Multiple algorithms have been proposed to build from a corpus the most representative vocabulary of subwords, including Byte-Pair Encoding (BPE)~\cite{sennrich2016neural}, WordPiece~\cite{schuster2012japanese} or Unigram~\cite{kudo2018subword}.
BPE was initially developed as a compression algorithm~\cite{gage1994new} before being applied to text as a tokenization method.
The algorithm relies on creating new subword tokens by iteratively merging the most recurring pairs of successive tokens in a corpus until a chosen vocabulary size is reached. 
In the following, we call \textit{atomic elements} the tokens from the initial vocabulary and \textit{supertokens} the tokens added through BPE.

Some recent MIR studies have applied these algorithms to symbolic music~\cite{kumar2023words}.
BPE was first implemented to shorten token sequences~\cite{liu2022symphony}. 
\citet{fradet2023byte} specifically analyzed BPE for MIDI generation purposes and showed that the learned embedding spaces are more structured. 
However, when applied to piano analysis tasks, a BPE
with 4 times the initial vocabulary size
does not seem to show any downstream improvement in model performance~\cite{zhang2023symbolic}. 
In contrast, \citet{park2024mel2word} focus on specifically applying BPE on monophonic tunes using a pitch/duration-only representation and show that BPE enables the retrieval of style-specific motifs.

To date, research on BPE for symbolic music has focused on its evaluation on generation or global sequence classification tasks. Its behaviour has not been analyzed in depth, in particular when applied to various instrumentations.
This work specifically focuses on these issues, with a descriptive analysis of BPE vocabularies followed by a quantitative evaluation of BPE on monophonic and polyphonic music on a musical phrase segmentation task.

Our experiments rely on the MidiTok package~\cite{fradet2021miditok} to handle the tokenization process and the HuggingFace library~\cite{wolf2020transformers} implementing Transformer models. 
We publicly release the datasets and source code, which are available at \url{http://algomus.fr/code/}.

\section{Analyzing music BPE}
\label{sec:analyzing}

    \pgfplotsset{%
  every axis legend/.append style ={%
    anchor = north west,%
    at = {(0.02,0.97)}%
  },
  /pgf/number format/.cd,
  use comma,
  1000 sep = {\,},
  min exponent for 1000 sep = 4,
}

\definecolor{tab-1}{HTML}{1f77b4} %
\definecolor{tab-2}{HTML}{ff7f0e} %
\definecolor{tab-3}{HTML}{2ca02c} %
\definecolor{tab-4}{HTML}{d62728} %
\definecolor{tab-5}{HTML}{9467bd} %
\definecolor{tab-6}{HTML}{8c564b} %
\definecolor{tab-7}{HTML}{e377c2} %
\definecolor{tab-8}{HTML}{7f7f7f} %
\definecolor{tab-9}{HTML}{bcbd22} %
\definecolor{tab-10}{HTML}{17becf} %

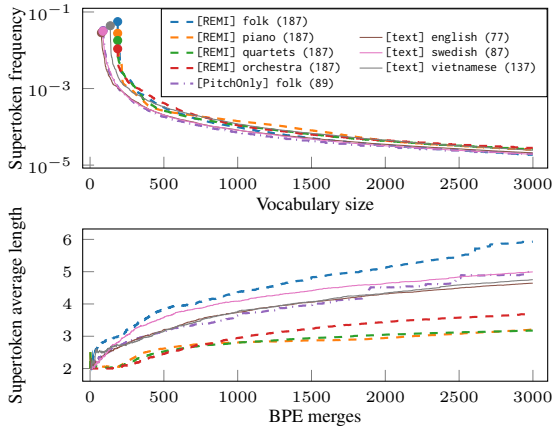
\begin{figure}[t]
    \centering
    \begin{tikzpicture}[remember picture]
      \begin{axis}[
          ymode=log,
          ylabel={Supertoken frequency},
          xlabel={Vocabulary size},
          xtick={0,500,1000,1500,2000,2500,3000},
          tick pos=left,
          xmin=-50, xmax=3100,
          no markers,
          cycle list name=color list,
          width=\columnwidth,height=3.7cm,
          x tick label style={
              rotate=0,
              anchor=north,
              font=\tiny
            },
          y tick label style={
              rotate=0,
              font=\tiny
            },
          y label style={
            at={(axis description cs:.1,.5)},anchor=south,
            font=\scriptsize
          },
          x label style={
            at={(axis description cs:.5,-.1)},anchor=south,
            font=\scriptsize
          },
          legend style={
              nodes={scale=.45, transform shape},
              font=\ttfamily,
              legend cell align={left},
              at={(1,1)},
              anchor=north east,
              legend image post style={xscale=0.5},
              /tikz/every even column/.append style={column sep=0.2cm}
          }, 
          legend columns=5, 
          transpose legend,
      ]
        \addplot[tab-1, dashed, thick] table[x=vocab_size,y=frq,col sep=comma]{data/music_vs_text_bpe/[REMI]_folk-supertoken_frq.csv};
        \addlegendentry{[REMI] folk (187)}
          
          \addplot[tab-2, dashed, thick] table[x=vocab_size,y=frq,col sep=comma]{data/music_vs_text_bpe/[REMI]_piano-supertoken_frq.csv};
          \addlegendentry{[REMI] piano (187)}
          
          \addplot[tab-3, dashed, thick] table[x=vocab_size,y=frq,col sep=comma]{data/music_vs_text_bpe/[REMI]_quartets-supertoken_frq.csv};
          \addlegendentry{[REMI] quartets (187)}
          
          \addplot[tab-4, dashed, thick] table[x=vocab_size,y=frq,col sep=comma]{data/music_vs_text_bpe/[REMI]_orchestra-supertoken_frq.csv};
          \addlegendentry{[REMI] orchestra (187)}
          
          \addplot[tab-5, dash dot, thick] table[x=vocab_size,y=frq,col sep=comma]{data/music_vs_text_bpe/[PitchOnly]_folk-supertoken_frq.csv};
          \addlegendentry{[PitchOnly] folk (89)}
          
          \addlegendimage{empty legend}
          \addlegendentry{}

          \addplot[tab-6] table[x=vocab_size,y=frq,col sep=comma]{data/music_vs_text_bpe/[text]_english-supertoken_frq.csv};
          \addlegendentry{[text] english (77)}
          
          \addplot[tab-7] table[x=vocab_size,y=frq,col sep=comma]{data/music_vs_text_bpe/[text]_swedish-supertoken_frq.csv};
          \addlegendentry{[text] swedish (87)}
          
          \addplot[tab-8] table[x=vocab_size,y=frq,col sep=comma]{data/music_vs_text_bpe/[text]_vietnamese-supertoken_frq.csv};
          \addlegendentry{[text] vietnamese (137)}

          \addplot[tab-1,only marks, mark size=1.5pt] table[x=init_vocab_size,y=init_frq,col sep=comma]{data/music_vs_text_bpe/[REMI]_folk-vocab_init_size.csv};
          \addplot[tab-2,only marks, mark size=1.5pt] table[x=init_vocab_size,y=init_frq,col sep=comma]{data/music_vs_text_bpe/[REMI]_piano-vocab_init_size.csv};
          \addplot[tab-3,only marks, mark size=1.5pt] table[x=init_vocab_size,y=init_frq,col sep=comma]{data/music_vs_text_bpe/[REMI]_quartets-vocab_init_size.csv};
          \addplot[tab-4,only marks, mark size=1.5pt] table[x=init_vocab_size,y=init_frq,col sep=comma]{data/music_vs_text_bpe/[REMI]_orchestra-vocab_init_size.csv};
          \addplot[tab-5,only marks, mark size=1.5pt] table[x=init_vocab_size,y=init_frq,col sep=comma]{data/music_vs_text_bpe/[PitchOnly]_folk-vocab_init_size.csv};
          \addplot[tab-6,only marks, mark size=1.5pt] table[x=init_vocab_size,y=init_frq,col sep=comma]{data/music_vs_text_bpe/[text]_english-vocab_init_size.csv};
          \addplot[tab-7,only marks, mark size=1.5pt] table[x=init_vocab_size,y=init_frq,col sep=comma]{data/music_vs_text_bpe/[text]_swedish-vocab_init_size.csv};
          \addplot[tab-8,only marks, mark size=1.5pt] table[x=init_vocab_size,y=init_frq,col sep=comma]{data/music_vs_text_bpe/[text]_vietnamese-vocab_init_size.csv};

      \end{axis}
    \end{tikzpicture} 

    \vspace{-1em}
      
    \begin{tikzpicture} 

        \begin{axis}[
            ylabel={Supertoken average length}, ylabel style={align=center},
            xlabel={BPE merges},
            xtick={0,500,1000,1500,2000,2500,3000},
            ytick={2,3,4,5,6},
            tick pos=left,
            xmin=-50, xmax=3100,
            no markers,
            cycle list name=color list,
            width=\columnwidth,height=3.6cm,
            x tick label style={
                rotate=0,
                anchor=north,
                font=\tiny
              },
            y tick label style={
                rotate=0,
                font=\tiny
              },
            y label style={
              at={(axis description cs:.1,.5)},anchor=south,
              font=\scriptsize
            },
            x label style={
              at={(axis description cs:.5,-.1)},anchor=south,
              font=\scriptsize
            },
        ]
          \addplot[tab-1, dashed, thick] table[x=bpe_merges,y=length,col sep=comma]{data/music_vs_text_bpe/[REMI]_folk-supertoken_length.csv};
            
          \addplot[tab-2, dashed, thick] table[x=bpe_merges,y=length,col sep=comma]{data/music_vs_text_bpe/[REMI]_piano-supertoken_length.csv};
          
          \addplot[tab-3, dashed, thick] table[x=bpe_merges,y=length,col sep=comma]{data/music_vs_text_bpe/[REMI]_quartets-supertoken_length.csv};
          
          \addplot[tab-4, dashed, thick] table[x=bpe_merges,y=length,col sep=comma]{data/music_vs_text_bpe/[REMI]_orchestra-supertoken_length.csv};
          
          \addplot[tab-5, dash dot, thick] table[x=bpe_merges,y=length,col sep=comma]{data/music_vs_text_bpe/[PitchOnly]_folk-supertoken_length.csv};
            
          \addplot[tab-6] table[x=bpe_merges,y=length,col sep=comma]{data/music_vs_text_bpe/[text]_english-supertoken_length.csv};
          
          \addplot[tab-7] table[x=bpe_merges,y=length,col sep=comma]{data/music_vs_text_bpe/[text]_swedish-supertoken_length.csv};
          
          \addplot[tab-8] table[x=bpe_merges,y=length,col sep=comma]{data/music_vs_text_bpe/[text]_vietnamese-supertoken_length.csv};
  
        \end{axis}
        \end{tikzpicture} 
    \caption{
        \textit{(Top)} Frequency of the created supertokens through the vocab size increasing with the BPE steps, for different styles of music and multilingual text data. \\
        \textit{(Bottom)} Average length of already created supertokens through BPE iterations for musical and text data. 
        The initial vocabulary size of each tokenization is indicated.
    }
    \label{fig:bpe_text_music}
\end{figure}

In this section, we present analyses of the vocabulary produced by Byte-Pair Encoding when applied to text and music. 
We first analyse supertokens induced by various instrumentations as well as their relation to high-level or abstract musical features.

\subsection{Comparing text and music BPEs}

Musical notes are often compared to text at the level of characters~\cite{hirata2022formal}.
Deep learning models have been shown to be more efficient when dealing with characters grouped into (sub)words~\cite{shapiro2018bpe,tay2021charformer}.
Therefore, we study the BPE results when processed, on text and music, in order to observe common or distinctive operating regime on such data with various languages and instrumentations.
Text data includes alphabetic\footnote{Experiments have also been conducted on syllabic (Japanese) and logographic (Chinese, Korean) languages, that show major differences due to the different nature of the atomic elements of their initial vocabulary.}
languages from various regions, extracted from the XLNI dataset~\cite{conneau2018xnli} on which we run BPE on 100k premises.
For music, we compare monophonic folk tunes, classical piano, string quartet, and orchestral corpora with similar sizes and tokenize these datasets using REMI~\cite{huang2020pop} from which \texttt{Velocity} tokens are removed.

We first study the occurrence frequency of the newly created supertoken within the corpus, at each step of the training (Figure~\ref{fig:bpe_text_music}, top). 
To make the corpora and vocabularies comparable, supertoken frequencies are normalized by the initial corpus length, and the BPE iterations are aligned with the resulting vocabulary size.
Interestingly, the vocabularies obtained on music or text through BPE do not show major differences with respect to the decay rate or the order of magnitude of the frequencies.

We also compute the mean length of the supertokens through the BPE steps (Figure~\ref{fig:bpe_text_music}, bottom). 
The evolution of supertoken length differs between text and music, depending on the instrumentation.
While monophonic supertokens are generally longer than polyphonic ones, orchestra supertokens surprisingly appear to be longer than piano or string quartet ones.
An in-depth study of the constructed vocabulary shows that the orchestral vocabulary predominantly consists of "harmonic" supertokens formed of simultaneous notes. In contrast, piano and string quartet vocabularies include both simultaneous and consecutive notes.
This difference causes BPE to struggle to build long piano or string quartet supertokens. On a separated experiment, we observed that it takes over 10 times more steps on a piano corpus to get an average length comparable to that of the vocabulary obtained on the monophonic corpus.
Moreover, when considering an alphabet which only keeps pitch tokens, we show that monophonic supertoken lengths have a regime closer to that of text for this range of BPE merges (Figure~\ref{fig:bpe_text_music}, "PitchOnly" curve), while polyphonic curves still stand out.
We can thus posit that the differences between the music and text curves might be due to simultaneity and timing information, which are inherent to music.

\subsection{Musical content carried by supertokens}
\label{sec:musical_content}

So far, we have drawn a broad characterisation of the BPE vocabularies, let us now zoom in and try to delineate which supertokens are present in a specific context. 
Borrowed from text, the terms ``musical phrase'' or ``musical sentence''~\cite{nattiez1990music} denote a part of the music which can give the impression of a complete statement by its own.
The TAVERN dataset~\cite{devaney2015theme} include such phrase annotations.

Using a Structured~\cite{hadjeres2021piano} tokenization with pitches encoded as intervals~\cite{kermarec2022improving} 
we analyzed the segmentation induced on the sequences by a 1024-merge BPE. 
This tokenization allows taking advantage of both Structured's relative encoding of rhythm with time-shifts and the relative encoding of pitches through intervals.
A first observation is that only 4.2\% of the supertokens among the tokens of the sequences do overlap phrases. 
In contrast, randomly splitting the piece into the same number of chunks as BPE segmentation results in 71\% overlap ratio, indicating that supertokens are unlikely to span across phrase boundaries.

\begin{figure}[t]
    \centering
    \small
    \texttt{\scriptsize <Duration(4), TShift(4), Horizontal\_PitchInterval(+5)>} \\[3pt]
    \includegraphics[height=4.4em]{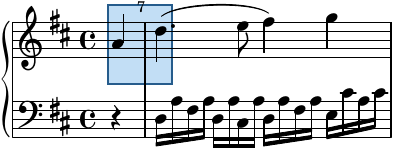}
    \hspace{7pt}
    \includegraphics[height=4.4em]{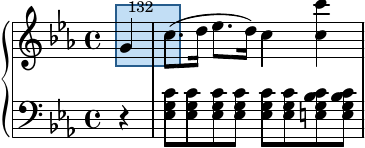}
    
    \vspace{1em}
    \texttt{\scriptsize
    <Duration(1), TShift(1), Horizontal\_PitchInterval(-5), \\ 
    Duration(1), TShift(1), Horizontal\_PitchInterval(-3), \\
    Duration(1), TShift(1), Horizontal\_PitchInterval(-4)>} \\[3pt]
    \raisebox{-4pt}{\includegraphics[height=5.3em]{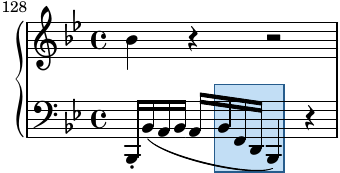}}
    \hspace{2pt}
    \includegraphics[height=4.8em]{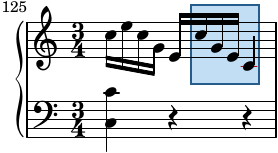}
    
    \caption{
        \textit{(Top)} First most common start-of-phrase supertoken from Mozart's K.25 and Beethoven's WoO.68. \\ %
        \textit{(Bottom)} 9-long common ending supertoken (10th most common) from Beethoven's WoO.73 and Mozart's K.179. %
        The tokenization is Structured + intervals.
    }
    \label{fig:end_of_phrase_supertokens}
\end{figure}

We then analysed the supertokens occurring at the beginning and end of musical phrases. 
In particular, our chosen tokenization allows this analysis to be key signature-independent and bar position-independent.
The most recurrent start-of-phrase supertoken appears to be a melodic rising perfect fourth (Figure~\ref{fig:end_of_phrase_supertokens}, top), which follows musicology studies~\cite[p.145]{meyer1973explaining}: 
``\textit{an upbeat interval of a perfect fourth, moving to the tonic [...] may be understood as a rhythmic-harmonic event emphasizing the tonic on which the melody proper begins.}''
Most represented end-of-phrase supertokens include descending arpeggio patterns on the tonic chord (Figure~\ref{fig:end_of_phrase_supertokens}). 
This also verifies some musicological observations~\cite{huron1996melodic}: 
``\textit{Melodic passages tend to exhibit an arch shape where the overall pitch contour rises and then falls over the course of a phrase or an entire melody}''.
Therefore, similar to how BPE can capture syntactic rules in text, we observe that musical supertokens also convey high-level musical information.

\section{Evaluating BPE on musical phrase segmentation}
\label{sec:experiments}

BPE applied to MIDI-derived tokenization has been mainly evaluated through classification tasks with composer classification, on a general multi-track dataset~\cite{fradet2023byte} or specifically piano music~\cite{zhang2023symbolic}.
Inspired by sentence segmentation tasks in NLP~\cite{read2012sentence} and given our preliminary results showing that supertokens can play a role in musical phrase boundaries, we aim to quantitatively evaluate BPE on a task of musical phrase segmentation for monophonic and polyphonic datasets.

\subsection{Musical phrase segmentation}

We consider a \textit{musical phrase segmentation task}, where a model is trained to tag each token of a sequence as being a start-of-phrase or not~\cite{guan2018melodic}. %
For BPE sequences, if a start-of-phrase occurs within a supertoken, the whole supertoken is annotated as being a start-of-phrase.

We first performed this task on the MTC dataset~\cite{van2014meertens} composed of monophonic Dutch folk tunes and including phrase annotations. 
The MTC dataset contains 100 times more phrase annotations than TAVERN.
Moreover, the nature of classical-style musical phrases, generally based on cadences~\cite{spencer1994practical}, may differ from folk music phrases, based on melodic contours~\cite{huron1996melodic}. 
Therefore, for a fairer comparison, we discard TAVERN as our polyphonic dataset and we build and release a 
synthetic
dataset of folk music piano arrangements from the MTC dataset generated by the AccoMontage model~\cite{zhao2021accomontage} aligned with the original phrase annotations. 
We tokenize both datasets using REMI~\cite{huang2020pop} and remove the \texttt{Velocity} tokens, for simplicity.

Note that the non-BPE dataset is by design more unbalanced than the BPE one.
In the polyphonic setting, the proportion of start-of-phrases increases from 1.2\% in the whole dataset to 3.3\% after 128k merges, respectively from 2\% to 27\% in the monophonic dataset. 

\subsection{Experiments}

We trained a 2-layer Transformer encoder-only model with 8 heads per layer
and a common embedding size between BPE and non-BPE vocabularies on each dataset.
We evaluate each model on 3 different splits of the datasets, using the F1-score of the start of phrase label prediction.
As our experiments focus on representation impact, we chose to have light models rather than ones achieving optimal performance.

    \pgfplotsset{%
  every axis legend/.append style ={%
    anchor = north west,%
    at = {(0.02,0.97)}%
  },
  /pgf/number format/.cd,
  use comma,
  1000 sep = {\,},
  min exponent for 1000 sep = 4,
}

\definecolor{tab-1}{HTML}{1f77b4} %
\definecolor{tab-2}{HTML}{ff7f0e} %
\definecolor{tab-3}{HTML}{2ca02c} %
\definecolor{tab-4}{HTML}{d62728} %
\definecolor{tab-5}{HTML}{9467bd} %
\definecolor{tab-6}{HTML}{8c564b} %
\definecolor{tab-7}{HTML}{e377c2} %
\definecolor{tab-8}{HTML}{7f7f7f} %
\definecolor{tab-9}{HTML}{bcbd22} %
\definecolor{tab-10}{HTML}{17becf} %

\newcommand\tinier{\fontsize{4pt}{6pt}\selectfont}
\newcommand\tinierless{\fontsize{5pt}{7pt}\selectfont}
\newcommand\titlefontsize{\fontsize{8pt}{8pt}\selectfont}

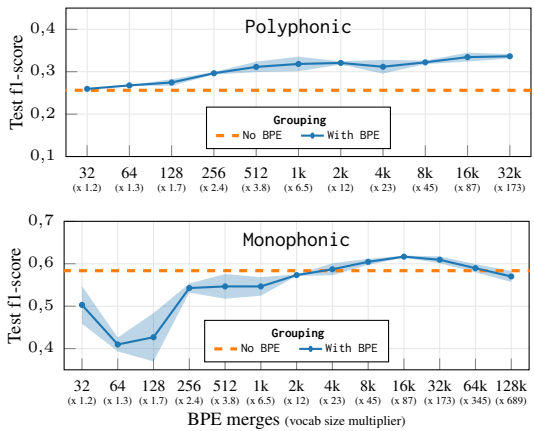
\begin{figure}[t]
    \centering
    \begin{tikzpicture}
        \begin{axis}[
            ylabel={Test f1-score},
            xlabel={},
            xtick={0,1,2,3,4,5,6,7,8,9,10},
            table/col sep = comma,
            xticklabels from table={data/perfo_mono_poly/perfo_monophonic.csv}{label},
            xticklabel style={align=center},
            ymin=0.1, ymax=0.45,
            xmin=-0.5, xmax=10.5,
            tick pos=left,
            cycle list name=color list,
            width=\columnwidth,height=3.55cm,
            title style={
                at={(0.5,0.89)},anchor=north,
            },
            title={\titlefontsize \ttfamily Polyphonic},
            x tick label style={
                rotate=0,
                anchor=north,
                font=\tiny
              },
            y tick label style={
                rotate=0,
                font=\tiny
              },
            y label style={
              at={(axis description cs:.13,.5)},anchor=south,
              font=\scriptsize
            },
            x label style={
              at={(axis description cs:.5,-.2)},anchor=south,
              font=\scriptsize
            },
            legend style={
                nodes={scale=.45, transform shape},
                font=\ttfamily,
                legend cell align={left},
                at={(0.5,0.05)},
                anchor=south,
                legend image post style={xscale=0.5},
                /tikz/every even column/.append style={column sep=0.2cm}
            }, 
            legend columns=2, 
            grid=both,
            grid style={line width=.1pt, draw=gray!20},
        ]
        \addlegendimage{empty legend} %
        \addlegendentry{\makebox[0pt][l]{\hspace{.9cm}\textbf{Grouping}}}
        
        \addlegendimage{empty legend}
        \addlegendentry{}

        \addplot[tab-2, very thick, dashed] table[x=x,y=zero_perf,col sep=comma]{data/perfo_mono_poly/perfo_polyphonic_zero.csv};
        \addlegendentry{No BPE}

        \addplot[tab-1, thick, mark=*, mark size=.75pt] table[x=x,y=f1_score_mean,col sep=comma]{data/perfo_mono_poly/perfo_polyphonic.csv};
        \addplot [name path=upper,draw=none] table[x=x,y expr=\thisrow{f1_score_mean}+\thisrow{f1_score_ci},col sep=comma] {data/perfo_mono_poly/perfo_polyphonic.csv};
        \addplot [name path=lower,draw=none] table[x=x,y expr=\thisrow{f1_score_mean}-\thisrow{f1_score_ci},col sep=comma] {data/perfo_mono_poly/perfo_polyphonic.csv};
        \addplot [fill=tab-1,fill opacity=0.3] fill between[of=upper and lower];
        \addlegendentry{With BPE}
        
    \end{axis}
    \end{tikzpicture}

    \begin{tikzpicture}
        \begin{axis}[
            ylabel={Test f1-score},
            xlabel={BPE merges {\tinierless (vocab size multiplier)}},
            xtick={0,1,2,3,4,5,6,7,8,9,10,11,12},
            table/col sep = comma,
            xticklabels from table={data/perfo_mono_poly/perfo_monophonic.csv}{label},
            xticklabel style={align=center},
            ymin=0.35, ymax=0.7,
            xmin=-0.5, xmax=12.5,
            tick pos=left,
            cycle list name=color list,
            width=\columnwidth,height=3.55cm,
            title style={
                at={(0.5,0.89)},anchor=north,
            },
            title={\titlefontsize \ttfamily Monophonic},
            x tick label style={
                rotate=0,
                anchor=north,
                font=\tiny
              },
            y tick label style={
                rotate=0,
                font=\tiny
              },
            y label style={
              at={(axis description cs:.13,.5)},anchor=south,
              font=\scriptsize
            },
            x label style={
              at={(axis description cs:.5,-.2)},anchor=south,
              font=\scriptsize
            },
            legend style={
                nodes={scale=.45, transform shape},
                font=\ttfamily,
                legend cell align={left},
                at={(0.5,0.05)},
                anchor=south,
                legend image post style={xscale=0.5},
                /tikz/every even column/.append style={column sep=0.2cm}
            }, 
            legend columns=2, 
            grid=both,
            grid style={line width=.1pt, draw=gray!20},
        ]
        \addlegendimage{empty legend} %
        \addlegendentry{\makebox[0pt][l]{\hspace{.9cm}\textbf{Grouping}}}
        
        \addlegendimage{empty legend}
        \addlegendentry{}

        \addplot[tab-2, very thick, dashed] table[x=x,y=zero_perf,col sep=comma]{data/perfo_mono_poly/perfo_monophonic_zero.csv};
        \addlegendentry{No BPE}

        \addplot [tab-1, thick, mark=*, mark size=.75pt] table[x=x,y=f1_score_mean,col sep=comma]{data/perfo_mono_poly/perfo_monophonic.csv};
        \addplot [name path=upper,draw=none] table[x=x,y expr=\thisrow{f1_score_mean}+\thisrow{f1_score_ci},col sep=comma] {data/perfo_mono_poly/perfo_monophonic.csv};
        \addplot [name path=lower,draw=none] table[x=x,y expr=\thisrow{f1_score_mean}-\thisrow{f1_score_ci},col sep=comma] {data/perfo_mono_poly/perfo_monophonic.csv};
        \addplot [fill=tab-1,fill opacity=0.3] fill between[of=upper and lower];
        \addlegendentry{With BPE}
        
    \end{axis}
    \end{tikzpicture}
    
    \caption{
         f1-score for start-of-phrase classification on the polyphonic \textit{(top)} and monophonic dataset \textit{(bottom)}.
    }
    \label{fig:perfo}
\end{figure}

The polyphonic setting of our experiment seems to indicate that BPE can have an impact on performance. Indeed, unlike \citet{zhang2023symbolic} also focusing on piano music, who demonstrated on a sequence global classification task that a BPE (with the initial vocabulary size $\times 4$) does not result in significant improvements, we see on this local classification task that the performance increases with the number of merges (Figure~\ref{fig:perfo}, top).

Our results on the monophonic dataset show even that BPE with too few number of merges can degrade the performance (Figure~\ref{fig:perfo}, bottom). This surprising behavior also occurs in NLP tasks, where character-based models can outperform subword-based models~\cite{chung2016character}.

Figure~\ref{fig:pitch_supertokens} describes the "melodic" content of the supertokens created along BPE steps. An analysis of supertokens reveals that early merges tend to produce \textit{structural} supertokens, such as combinations of \texttt{Bar} and \texttt{Beat} (Figure~\ref{fig:pitch_supertokens} gray area: proportion of created supertokens with 0 \texttt{<Pitch>} atomic element), while melodic patterns emerge later, and at different rates for monophonic and polyphonic datasets. At 128 merges (Figure 4, dashed line), 26\% of monophonic supertokens do not include any \texttt{<Pitch>} atomic element (gray area) while this ratio is only 9\% for polyphonic and 7\% contain 2 \texttt{<Pitch>} atomic element (green area). Fewer melodic patterns, which are more likely to indicate phrase boundaries in monophonic tunes~\cite{huron1996melodic}, may explain why the BPE model performs better only after a certain number of merges.

In the monophonic dataset we also see that, after too many merges, the model performance drops. An analysis of the supertoken length shows that, after 128k merges, monophonic supertokens are on average 38.6-long (compared to 8.4 for polyphonic ones). Indeed, the smaller size of the monophonic dataset ($3\times$ smaller than the polyphonic one) leads late steps supertokens to capture long but rare patterns that might be less relevant for this task of phrase segmentation.

    \pgfplotsset{%
  every axis legend/.append style ={%
    anchor = north west,%
    at = {(0.02,0.97)}%
  },
  /pgf/number format/.cd,
  use comma,
  1000 sep = {\,},
  min exponent for 1000 sep = 4,
}

\definecolor{ygb-1}{HTML}{999999} %
\definecolor{ygb-2}{HTML}{ffffd9} %
\definecolor{ygb-3}{HTML}{c6e9b4} %
\definecolor{ygb-4}{HTML}{40b5c4} %
\definecolor{ygb-5}{HTML}{225da8} %
\definecolor{ygb-6}{HTML}{081d58} %

\pgfplotsset{
    custom area legend/.style={
        legend image code/.code={%
            \path[#1,draw=none](0cm,-.05cm)rectangle(.5cm,.05cm);%
        }
    }
}

\newcommand\titlefontsizesmall{\fontsize{6pt}{6pt}\selectfont}

\begin{figure}[t]
    \centering
    \begin{tikzpicture}[remember picture]
        \begin{groupplot}[
            group style={
              group size=2 by 1,
              horizontal sep=.3cm,
            },
            width=\columnwidth,
          ]
        \nextgroupplot[
            ylabel={{Supertokens\\with $n$ \texttt{<Pitch>}}},
            ylabel style={align=center},
            xlabel={Merges},
            xtick={0,500,1000,1500,2000,2500,3000},
            ytick={0,0.25,0.5,0.75,1},
            yticklabels={0\%,25\%,50\%,75\%,100\%},
            tick pos=left,
            xmin=0, xmax=2048,
            ymin=0, ymax=1,
            no markers,
            cycle list name=color list,
            width=.56\columnwidth,height=3.4cm,
            x tick label style={
                rotate=0,
                anchor=north,
                font=\tiny
              },
            y tick label style={
                rotate=0,
                font=\tiny
              },
            y label style={
              at={(axis description cs:0.22,.5)},anchor=south,
              font=\scriptsize
            },
            x label style={
              at={(axis description cs:.5,-.1)},anchor=south,
              font=\scriptsize
            },
            legend style={
                nodes={scale=.45, transform shape},
                font=\ttfamily\footnotesize,
                legend cell align={left},
                at={(.96, 0.4)},
                anchor=east,
                legend image post style={xscale=0.5},
                /tikz/every even column/.append style={column sep=0.2cm}
            }, 
            legend columns=1, 
            table/col sep = comma,
            axis on top
        ]
            \addlegendimage{custom area legend,fill=ygb-5}
            \addlegendentry{4 <Pitch>}
            \addlegendimage{custom area legend,fill=ygb-4}
            \addlegendentry{3 <Pitch>}
            \addlegendimage{custom area legend,fill=ygb-3}
            \addlegendentry{2 <Pitch>}
            \addlegendimage{custom area legend,fill=ygb-2}
            \addlegendentry{1 <Pitch>}
            \addlegendimage{custom area legend,fill=ygb-1}
            \addlegendentry{0 <Pitch>}

          \path[name path=axis] (axis cs:0,0) -- (axis cs:3000,0);
          \addplot [name path=pitch0, opacity=0] table[x=x,y=p0]{data/ratio_pitch_supertokens/ratio_pitch_supertokens_poly.csv};
          \addplot [thick, fill=ygb-1] fill between[of=axis and pitch0];

          \addplot [name path=pitch1, opacity=0] table[x=x, y expr=\thisrow{p0}+\thisrow{p1}]{data/ratio_pitch_supertokens/ratio_pitch_supertokens_poly.csv};
          \addplot [thick, fill=ygb-2] fill between[of=pitch0 and pitch1];
          
          \addplot [name path=pitch2, opacity=0] table[x=x, y expr=\thisrow{p0}+\thisrow{p1}+\thisrow{p2}]{data/ratio_pitch_supertokens/ratio_pitch_supertokens_poly.csv};
          \addplot [thick, fill=ygb-3] fill between[of=pitch1 and pitch2];
          
          \addplot [name path=pitch3, opacity=0] table[x=x, y expr=\thisrow{p0}+\thisrow{p1}+\thisrow{p2}+\thisrow{p3}]{data/ratio_pitch_supertokens/ratio_pitch_supertokens_poly.csv};
          \addplot [thick, fill=ygb-4] fill between[of=pitch2 and pitch3];
          
          \addplot [name path=pitch4, opacity=0] table[x=x, y expr=\thisrow{p0}+\thisrow{p1}+\thisrow{p2}+\thisrow{p3}+\thisrow{p4}]{data/ratio_pitch_supertokens/ratio_pitch_supertokens_poly.csv};
          \addplot [thick, fill=ygb-5] fill between[of=pitch3 and pitch4];
            
          \addplot[tab-2, very thick, dashed] coordinates {(128, 0) (128, 1)} ;

          \node[draw=gray!30, fill=white, opacity=.6, text opacity=1, rectangle, rounded corners=2, inner sep=2pt] at (110, 87) {\titlefontsizesmall \ttfamily Polyphonic};
    \nextgroupplot[
            ylabel style={align=center},
            xlabel={Merges},
            xtick={0,500,1000,1500,2000,2500,3000},
            ytick={0,0.25,0.5,0.75,1},
            yticklabels={},
            tick pos=left,
            xmin=0, xmax=2048,
            ymin=0, ymax=1,
            no markers,
            cycle list name=color list,
            width=.58\columnwidth,height=3.4cm,
            x tick label style={
                rotate=0,
                anchor=north,
                font=\tiny
              },
            x label style={
              at={(axis description cs:.5,-.1)},anchor=south,
              font=\scriptsize
            },
            table/col sep = comma,
            axis on top
        ]
          \path[name path=axis] (axis cs:0,0) -- (axis cs:3000,0);
          \addplot [name path=pitch0, opacity=0] table[x=x,y=p0]{data/ratio_pitch_supertokens/ratio_pitch_supertokens_mono.csv};
          \addplot [thick, fill=ygb-1] fill between[of=axis and pitch0];
          
          \addplot [name path=pitch1, opacity=0] table[x=x, y expr=\thisrow{p0}+\thisrow{p1}]{data/ratio_pitch_supertokens/ratio_pitch_supertokens_mono.csv};
          \addplot [thick, fill=ygb-2] fill between[of=pitch0 and pitch1];
          
          \addplot [name path=pitch2, opacity=0] table[x=x, y expr=\thisrow{p0}+\thisrow{p1}+\thisrow{p2}]{data/ratio_pitch_supertokens/ratio_pitch_supertokens_mono.csv};
          \addplot [thick, fill=ygb-3] fill between[of=pitch1 and pitch2];
          
          \addplot [name path=pitch3, opacity=0] table[x=x, y expr=\thisrow{p0}+\thisrow{p1}+\thisrow{p2}+\thisrow{p3}]{data/ratio_pitch_supertokens/ratio_pitch_supertokens_mono.csv};
          \addplot [thick, fill=ygb-4] fill between[of=pitch2 and pitch3];
          
          \addplot [name path=pitch4, opacity=0] table[x=x, y expr=\thisrow{p0}+\thisrow{p1}+\thisrow{p2}+\thisrow{p3}+\thisrow{p4}]{data/ratio_pitch_supertokens/ratio_pitch_supertokens_mono.csv};
          \addplot [thick, fill=ygb-5] fill between[of=pitch3 and pitch4];
          
          \addplot[tab-2, very thick, dashed] coordinates {(128, 0) (128, 1)} ;

          \node[draw=gray!30, fill=white, opacity=.6, text opacity=1, rectangle, rounded corners=2, inner sep=2pt] at (110, 87) {\titlefontsizesmall \ttfamily Monophonic};
            
        \end{groupplot}
      \end{tikzpicture} 
    \caption{
         Ratio of supertokens containing $n$ \texttt{<Pitch>} atomic elements in the vocabulary for each number of BPE merges.
    }
    \label{fig:pitch_supertokens}
\end{figure}
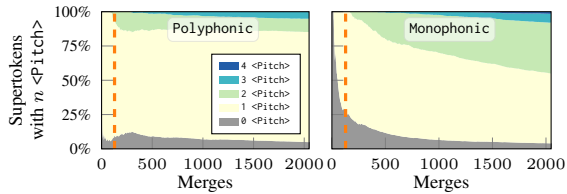

\section{Conclusion}
\label{sec:conclusion}

In this work, we show that Byte-Pair Encoding behaves differently depending on the type of music it is trained on.
With a descriptive approach, we highlight that the resulting vocabulary highly depends on the type of instrumentation, and supertokens can carry high-level musical content.
On a downstream task, we confirm the impact of instrumentation on the model performance and show that the number of BPE merges should be chosen carefully. For future work, we think the initial tokenization impact over BPE performance should be investigated.

\section*{Acknowledgments}
This work was supported by grant ANR-20-THIA-0014 program ``AI\_PhD@Lille''.
The authors would like to thank Zih-Syuan Lin for providing feedbacks on earlier versions of the paper.

\bibliography{references}

\begin{thebibliography}{31}
\providecommand{\natexlab}[1]{#1}

\bibitem[{Chung et~al.(2016)Chung, Cho, and Bengio}]{chung2016character}
Junyoung Chung, Kyunghyun Cho, and Yoshua Bengio. 2016.
\newblock \href {https://doi.org/10.18653/v1/P16-1160} {A character-level
  decoder without explicit segmentation for neural machine translation}.
\newblock In \emph{Proceedings of the 54th Annual Meeting of the Association
  for Computational Linguistics (Volume 1: Long Papers)}, pages 1693--1703,
  Berlin, Germany. Association for Computational Linguistics.

\bibitem[{Conneau et~al.(2018)Conneau, Rinott, Lample, Williams, Bowman,
  Schwenk, and Stoyanov}]{conneau2018xnli}
Alexis Conneau, Ruty Rinott, Guillaume Lample, Adina Williams, Samuel Bowman,
  Holger Schwenk, and Veselin Stoyanov. 2018.
\newblock \href {https://doi.org/10.18653/v1/D18-1269} {{XNLI}: Evaluating
  cross-lingual sentence representations}.
\newblock In \emph{Proceedings of the 2018 Conference on Empirical Methods in
  Natural Language Processing}, pages 2475--2485, Brussels, Belgium.
  Association for Computational Linguistics.

\bibitem[{Devaney et~al.(2015)Devaney, Arthur, Condit-Schultz, and
  Nisula}]{devaney2015theme}
Johanna Devaney, Claire Arthur, Nathaniel Condit-Schultz, and Kirsten Nisula.
  2015.
\newblock \href {https://doi.org/10.5281/zenodo.1417496} {Theme and variation
  encodings with roman numerals ({TAVERN}): {A} new data set for symbolic music
  analysis}.
\newblock In \emph{International Society for Music Information Retrieval
  Conference ({ISMIR})}.

\bibitem[{Fradet et~al.(2021)Fradet, Briot, Chhel, El~Fallah-Seghrouchni, and
  Gutowski}]{fradet2021miditok}
Nathan Fradet, Jean-Pierre Briot, Fabien Chhel, Amal El~Fallah-Seghrouchni, and
  Nicolas Gutowski. 2021.
\newblock \href {https://doi.org/10.48550/arXiv.2310.17202} {Midi{T}ok: A
  {P}ython package for {MIDI} file tokenization}.
\newblock In \emph{International Society for Music Information Retrieval
  Conference (ISMIR), Late-Breaking Demo Session}.

\bibitem[{Fradet et~al.(2023)Fradet, Gutowski, Chhel, and
  Briot}]{fradet2023byte}
Nathan Fradet, Nicolas Gutowski, Fabien Chhel, and Jean-Pierre Briot. 2023.
\newblock \href {https://doi.org/10.18653/v1/2023.emnlp-main.123} {Byte pair
  encoding for symbolic music}.
\newblock In \emph{Proceedings of the 2023 Conference on Empirical Methods in
  Natural Language Processing}, pages 2001--2020, Singapore. Association for
  Computational Linguistics.

\bibitem[{Gage(1994)}]{gage1994new}
Philip Gage. 1994.
\newblock \href {https://doi.org/10.5555/177910.177914} {A new algorithm for
  data compression}.
\newblock \emph{C~Users Journal}, 12(2):23--38.

\bibitem[{Guan et~al.(2018)Guan, Zhao, Qiu, Zhang, and Xia}]{guan2018melodic}
Yixing Guan, Jinyu Zhao, Yiqin Qiu, Zheng Zhang, and Gus Xia. 2018.
\newblock \href {https://arxiv.org/abs/1811.05688} {Melodic phrase segmentation
  by deep neural networks}.
\newblock \emph{Preprint}, arXiv:1811.05688.

\bibitem[{Hadjeres and Crestel(2021)}]{hadjeres2021piano}
Gaëtan Hadjeres and Léopold Crestel. 2021.
\newblock \href {https://arxiv.org/abs/2107.05944} {The piano inpainting
  application}.
\newblock \emph{Preprint}, arXiv:2107.05944.

\bibitem[{Hirata et~al.(2022)Hirata, Tojo, and Hamanaka}]{hirata2022formal}
Keiji Hirata, Satoshi Tojo, and Masatoshi Hamanaka. 2022.
\newblock \href {https://doi.org/10.1007/978-981-19-5166-4_3} {Music as formal
  language}.
\newblock In \emph{Music, Mathematics and Language: The New Horizon of
  Computational Musicology Opened by Information Science}, pages 51--78,
  Singapore. Springer Nature Singapore.

\bibitem[{Huang and Yang(2020)}]{huang2020pop}
Yu-Siang Huang and Yi-Hsuan Yang. 2020.
\newblock \href {https://doi.org/10.1145/3394171.3413671} {Pop music
  transformer: Beat-based modeling and generation of expressive pop piano
  compositions}.
\newblock In \emph{Proceedings of the 28th ACM International Conference on
  Multimedia}, MM '20, page 1180–1188, New York, NY, USA. Association for
  Computing Machinery.

\bibitem[{Huron et~al.(1996)}]{huron1996melodic}
David Huron et~al. 1996.
\newblock \href
  {http://www.music.mcgill.ca/~ich/classes/mumt621_20/papers/huron96melodic.pdf}
  {The melodic arch in western folksongs}.
\newblock \emph{Computing in Musicology}, 10:3--23.

\bibitem[{Jackendoff(2009)}]{jackendoff2009parallels}
Ray Jackendoff. 2009.
\newblock \href {http://www.jstor.org/stable/10.1525/mp.2009.26.3.195}
  {Parallels and nonparallels between language and music}.
\newblock \emph{Music Perception: An Interdisciplinary Journal},
  26(3):195--204.

\bibitem[{Kermarec et~al.(2022)Kermarec, Bigo, and
  Keller}]{kermarec2022improving}
Mathieu Kermarec, Louis Bigo, and Mikaela Keller. 2022.
\newblock \href {https://archives.ismir.net/ismir2022/latebreaking/000008.pdf}
  {Improving tokenization expressiveness with pitch intervals}.
\newblock In \emph{International Society for Music Information Retrieval
  Conference (ISMIR), Late-Breaking Demo Session}.

\bibitem[{Kudo(2018)}]{kudo2018subword}
Taku Kudo. 2018.
\newblock \href {https://doi.org/10.18653/v1/P18-1007} {Subword regularization:
  Improving neural network translation models with multiple subword
  candidates}.
\newblock In \emph{Proceedings of the 56th Annual Meeting of the Association
  for Computational Linguistics (Volume 1: Long Papers)}, pages 66--75,
  Melbourne, Australia. Association for Computational Linguistics.

\bibitem[{Kumar and Sarmento(2023)}]{kumar2023words}
Adarsh Kumar and Pedro Sarmento. 2023.
\newblock \href {https://arxiv.org/abs/2304.08953} {From words to music: A
  study of subword tokenization techniques in symbolic music generation}.
\newblock \emph{Preprint}, arXiv:2304.08953.

\bibitem[{Le et~al.(2024)Le, Bigo, Keller, and Herremans}]{le2024natural}
Dinh-Viet-Toan Le, Louis Bigo, Mikaela Keller, and Dorien Herremans. 2024.
\newblock \href {https://arxiv.org/abs/2402.17467} {Natural language processing
  methods for symbolic music generation and information retrieval: A survey}.
\newblock \emph{Preprint}, arXiv:2402.17467.

\bibitem[{Lerdahl(2013)}]{lerdahl2013musical}
Fred Lerdahl. 2013.
\newblock \href {https://doi.org/10.7551/mitpress/9548.003.0016} {{Musical
  Syntax and Its Relation to Linguistic Syntax}}.
\newblock In \emph{{Language, Music, and the Brain: A Mysterious
  Relationship}}. The MIT Press.

\bibitem[{Liu et~al.(2022)Liu, Dong, Cheng, Zhang, Li, Yu, and
  Sun}]{liu2022symphony}
Jiafeng Liu, Yuanliang Dong, Zehua Cheng, Xinran Zhang, Xiaobing Li, Feng Yu,
  and Maosong Sun. 2022.
\newblock \href {https://archives.ismir.net/ismir2022/paper/000066.pdf}
  {Symphony {Generation} with {Permutation} {Invariant} {Language} {Model}}.
\newblock In \emph{International Society for Music Information Retrieval
  Conference ({ISMIR})}.

\bibitem[{Meyer(1973)}]{meyer1973explaining}
Leonard~B. Meyer. 1973.
\newblock \href {https://doi.org/10.2307/jj.8501512} {\emph{Explaining Music:
  Essays and Explorations}}, {DGO} - {D}igital original edition.
\newblock University of California Press.

\bibitem[{Nattiez(1990)}]{nattiez1990music}
Jean-Jacques Nattiez. 1990.
\newblock \emph{Music and discourse: Toward a semiology of music}.
\newblock Princeton University Press.

\bibitem[{Park et~al.(2024)Park, Choi, Kim, and Nam}]{park2024mel2word}
Saebyul Park, Eunjin Choi, Jeounghoon Kim, and Juhan Nam. 2024.
\newblock \href {https://doi.org/10.1177/20592043231216254} {Mel2word: A
  text-based melody representation for symbolic music analysis}.
\newblock \emph{Music \& Science}, 7.

\bibitem[{Read et~al.(2012)Read, Dridan, Oepen, and Solberg}]{read2012sentence}
Jonathon Read, Rebecca Dridan, Stephan Oepen, and Lars~J{\o}rgen Solberg. 2012.
\newblock \href {https://aclanthology.org/C12-2096} {Sentence boundary
  detection: A long solved problem?}
\newblock In \emph{Proceedings of {COLING} 2012: Posters}, pages 985--994,
  Mumbai, India. The COLING 2012 Organizing Committee.

\bibitem[{Schuster and Nakajima(2012)}]{schuster2012japanese}
Mike Schuster and Kaisuke Nakajima. 2012.
\newblock \href {https://doi.org/10.1109/ICASSP.2012.6289079} {{J}apanese and
  {K}orean voice search}.
\newblock In \emph{2012 IEEE International Conference on Acoustics, Speech and
  Signal Processing (ICASSP)}, pages 5149--5152.

\bibitem[{Sennrich et~al.(2016)Sennrich, Haddow, and
  Birch}]{sennrich2016neural}
Rico Sennrich, Barry Haddow, and Alexandra Birch. 2016.
\newblock \href {https://doi.org/10.18653/v1/P16-1162} {Neural machine
  translation of rare words with subword units}.
\newblock In \emph{Proceedings of the 54th Annual Meeting of the Association
  for Computational Linguistics (Volume 1: Long Papers)}, pages 1715--1725,
  Berlin, Germany. Association for Computational Linguistics.

\bibitem[{Shapiro and Duh(2018)}]{shapiro2018bpe}
Pamela Shapiro and Kevin Duh. 2018.
\newblock \href {https://arxiv.org/abs/1809.01301} {{BPE} and {CharCNNs} for
  translation of morphology: A cross-lingual comparison and analysis}.
\newblock \emph{Preprint}, arXiv:1809.01301.

\bibitem[{Spencer and Temko(1994)}]{spencer1994practical}
Peter Spencer and Peter~M Temko. 1994.
\newblock \emph{A practical approach to the study of form in music}.
\newblock Waveland Press.

\bibitem[{Tay et~al.(2022)Tay, Tran, Ruder, Gupta, Chung, Bahri, Qin,
  Baumgartner, Yu, and Metzler}]{tay2021charformer}
Yi~Tay, Vinh~Q Tran, Sebastian Ruder, Jai Gupta, Hyung~Won Chung, Dara Bahri,
  Zhen Qin, Simon Baumgartner, Cong Yu, and Donald Metzler. 2022.
\newblock \href {https://openreview.net/forum?id=JtBRnrlOEFN} {Charformer: Fast
  character transformers via gradient-based subword tokenization}.
\newblock In \emph{International Conference on Learning Representations
  ({ICLR})}.

\bibitem[{Van~Kranenburg et~al.(2014)Van~Kranenburg, de~Bruin, Grijp, and
  Wiering}]{van2014meertens}
Peter Van~Kranenburg, MJ~de~Bruin, Louis~P Grijp, and Frans Wiering. 2014.
\newblock \href
  {https://www.liederenbank.nl/mtc/downloads/MeertensTuneCollections.pdf} {The
  {M}eertens tune collections}.
\newblock \emph{Meertens Online Reports}, 2014(1).

\bibitem[{Wolf et~al.(2020)}]{wolf2020transformers}
Thomas Wolf et~al. 2020.
\newblock \href {https://doi.org/10.18653/v1/2020.emnlp-demos.6} {Transformers:
  State-of-the-art natural language processing}.
\newblock In \emph{Proceedings of the 2020 Conference on Empirical Methods in
  Natural Language Processing: System Demonstrations}, pages 38--45, Online.
  Association for Computational Linguistics.

\bibitem[{Zhang et~al.(2023)Zhang, Karystinaios, Dixon, Widmer, and
  Cancino-Chac{\'o}n}]{zhang2023symbolic}
Huan Zhang, Emmanouil Karystinaios, Simon Dixon, Gerhard Widmer, and
  Carlos~Eduardo Cancino-Chac{\'o}n. 2023.
\newblock \href {10.5281/zenodo.10265420} {Symbolic music representations for
  classification tasks: A systematic evaluation}.
\newblock In \emph{International Society for Music Information Retrieval
  Conference ({ISMIR})}.

\bibitem[{Zhao and Xia(2021)}]{zhao2021accomontage}
Jingwei Zhao and Gus Xia. 2021.
\newblock \href {https://archives.ismir.net/ismir2021/paper/000104.pdf}
  {Accomontage: Accompaniment arrangement via phrase selection and style
  transfer}.
\newblock In \emph{Proceedings of the 22nd International Society for Music
  Information Retrieval Conference ({ISMIR} 2021)}, pages 833--840.

\end{thebibliography}

\end{document}